\documentclass[aps,prb,twocolumn,showpacs,amsmath,amssymb]{revtex4}
\usepackage{epsfig}
\usepackage{bm}
\begin{document}
\title{Spectral distribution and wavefunction of electrons emitted
  from a single particle source in the quantum Hall regime.}

\author{F. Battista and P. Samuelsson}
\affiliation{Division of
Mathematical Physics, Lund University, Box 118, S-221 00 Lund, Sweden}

\begin{abstract}
  We investigate theoretically a scheme for spectroscopy of electrons
  emitted by an on-demand single particle source. The total system,
  with an electron turnstile source and a single level quantum dot
  spectrometer, is implemented with edge states in a conductor in the
  quantum Hall regime. Employing a Floquet scattering approach, the
  source and the spectrometer are analyzed within a single theoretical 
  framework. The non-equilibrium distribution of the emitted electrons
  is analyzed via the direct current at the dot spectrometer. In the
  adiabatic and intermediate source frequency regimes, the
  distribution is found to be strongly peaked around the active
  resonant level of the turnstile. At high frequencies the
  distribution is split up into a set of fringes, resulting from the
  interplay of resonant transport through the turnstile and absorption
  or emission of individual Floquet quanta. For ideal source operation,
  with exactly one electron emitted per cycle, an expression for the
  single electron wavefunction is derived.
\end{abstract}

\pacs{72.10.-d, 73.23.-b, 73.43.-f}
\maketitle

\section{Introduction}
The last decade has shown an increasing interest in transport in the
integer quantum Hall regime, largely motivated by realizations of
electric analogs of fundamental quantum optics experiments. Conductors
in the quantum Hall regime provide the two key elements for electron
optics experiments: unidirectional edge states play the role of
electronic waveguides \cite{Halp,Butt} and quantum point contacts with
controllable transparency act as tunable electronic
beamsplitters. \cite{BS1,BS2,BS3} In their pioneering electron optics
experiment, Ji {\it et al}. \cite{MZI1} investigated an electronic
single particle, or Mach Zehnder, interferometer. \cite{Seelig} This
work was followed by a number of investigations, both experimental
\cite{MZI2,MZI3,MZI4,MZI5,MZI6,MZI7} and theoretical,
\cite{probe1,Vanessa,Forster,probe2,Sukh1,Chalk1,MZItheory1,Sukh,MZItheory2}
with the focus on the coherence and interaction properties of the
interferometer. Recently, following the proposal in
Ref. \onlinecite{2PItheory1}, a two-particle interferometer was
realized experimentally by Neder {\it et al}.. \cite{2PI} The
demonstration of two-particle interference provided a clear
experimental connection \cite{2PItheory2} between edge state transport
and quantum information
processing. \cite{Been03,Been03b,stace,Timeent1,Timeent2,tomo,Timeent3,giovannetti,Frustaglia,Bertoni}

Another important aspect of edge state transport, the high frequency
properties, was investigated in two key experiments.  Gabelli {\it et
  al.} \cite{Gabelli} analyzed the frequency dependent admittance of a
mesoscopic capacitor system. Good agreement was found with early
theoretical predictions, \cite{BPT} motivating additional
investigations focusing on the effects of electron-electron
interactions. \cite{Nigg,e-e1,e-e2,e-e3,e-e4,e-e5,e-e6}  In the experiment by Feve {\it et
  al}. \cite{Feve} a time controlled single particle source working at
gigahertz frequencies was realized.  It was demonstrated that a
mesoscopic capacitor coupled to an edge state can serve as a
time-periodic on-demand source, producing exactly one electron and one
hole per cycle.  The experiment was followed by a number of works
investigating the accuracy and coherence of the source
\cite{singem1,singem2,Mahe,Albert,Albert2,geraldine,parmalb} and also proposing novel
geometries with one or more on-demand sources as building blocks.
\cite{Theorypump1,Theorypump2,Theorypump3,Theorypump4} As an
interesting example, a scheme for time-bin entanglement generation
on-demand was proposed in Ref. \onlinecite{timebin}. Also other types
of edge state single electron sources were investigated, both
theoretically \cite{Battista} and experimentally. \cite{Leicht,Hohls}
Of particular importance for the present work is the nonlocal
electron-hole turnstile proposed by us, \cite{Battista} which during
ideal operation produces noiseless streams of electrons and holes
along spatially separated edges.

An additional important tool for investigations of edge state
transport was demonstrated recently by Altimiras {\it et
  al.}. \cite{pierre} They developed a method for a spectroscopic
analysis of the edge state distribution, by weakly coupling a quantum
dot with a single active level to the edge. In a series of works
\cite{pierre,pierrerelax,pierrerelax2} the energy relaxation and the
limitation of the electron-optics picture were investigated. Taken
together, the achievements in the field to date makes it both
experimentally accessible and fundamentally interesting to investigate
spectral properties of electronic states emitted from single
particle sources. A successful experiment would open up for a detailed
characterization of the state of the emitted particles. Moreover, for
a source emitting electrons well above Fermi energy, the modification
of the spectral properties of the particles propagating along the edge
is a sensitive tool for investigating electronic
interactions. \cite{Neuhahn2,relax1,Lunde}

In this work we perform a theoretical investigation of the
electron spectral properties by analyzing a combined single particle
source-spectral detector system implemented with edge states in a
multiterminal conductor, see Fig 1. As the single particle source we
consider the turnstile of Ref. \onlinecite{Battista}, although the
analysis can readily be extended to other
sources.\cite{Levitov,Feve,Vanevic,Timeent3} The distribution function of the
electrons emitted by the source is investigated via the direct current
flowing through the spectroscopic dot. We investigate the spectral
distribution for the three physically distinct turnstile frequency
regimes, adiabatic, intermediate and high, identified in
Ref. \onlinecite{Battista}. It is found that in the adiabatic and
intermediate regimes, the distribution is peaked around the energy of
the active resonance of the turnstile. At the cross-over to high
frequencies the peak splits up, developing fringes due to the Floquet
sidebands. At high frequencies a large number of features in the
spectral distribution appears, related to resonant transport through
higher lying turnstile levels. We discuss how these findings relate to
earlier work on time dependent transport in quantum dot and double
barrier systems. \cite{bruder,patkouvenI,patkouvenII,pedersen,Platero}
Moreover, we assess the robustness of our findings to
\textit{e.g.} rectification effects and stray capacitive couplings. In addition, in the ideal turnstile regime
we derive an expression for the wavefunction of single electrons
emitted from the turnstile, giving complete information about the
emitted state.

\section{Model}\label{modelsec}
\begin{figure}
\centerline{\psfig{figure=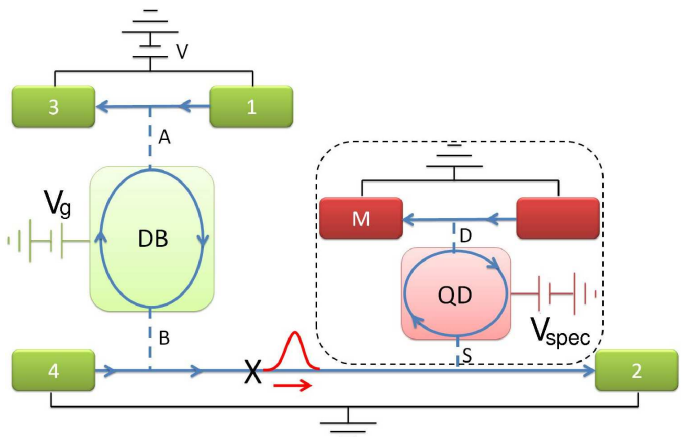,width=9cm}}
\caption{Schematic of the combined source-spectrometer system,
  implemented in a multiterminal conductor in the quantum Hall
  regime. Active spin-polarized edge states are shown with thick, blue
  lines, with arrows denoting the direction of propagation. The
  turnstile source (of Ref. \onlinecite{Battista}) is shown to the
  left. A bias difference $V$ is applied between terminals $1,3$ and
  $2,4$. Quantum point contacts $A$ and $B$ are driven by time
  periodic voltages. The double barrier (DB) region between $A$ and $B$ is
  capacitively coupled to a top gate (transparent green box), kept at
  a constant potential $V_g$. The emitted electron wavepacket is shown schematically in red at position $X$.
  The spectrometer (of Ref. \onlinecite{pierre}), shown inside the dashed box, consists
  of a quantum dot (QD), formed by static quantum point contacts $S$
  and $D$. The quantum dot, acting as an energy filter, has a single
  active level at energy $E_{spec}$, controlled by the voltage
  $V_{spec}$ applied to a top gate (red transparent box).  Current is
  measured at terminal $M$.}
\label{device}
\end{figure}

The combined source-spectrometer system is implemented in a
multiterminal conductor in the integer quantum Hall regime, see
Fig.~\ref{device}. Transport takes place along a single spin polarized
edge channel. 
The single particle source is the non-local electron-hole turnstile
proposed in Ref. \onlinecite{Battista}. Terminals $1$ and $3$ are
biased at $eV$ while terminals $2$ and $4$ are grounded. Electrons
scatter between edges at the two quantum point contacts $A$ and $B$,
driven by time periodic voltages $V_A(t)$ and $V_B(t)$ with a period
${\mathcal T}=2\pi/\omega$, $\pi$ out of phase. The time dependent
transparencies of the contacts are $T_A(t)$ and $T_B(t)$. Throughout
the paper, in the numerical analysis we model the contacts as saddle
point constrictions \cite{saddlepoint, saddlepoint2} with sinusoidal driving
potentials $V_A(t)=V_{A}^{\texttt{dc}}-V_{A}^{\texttt{ac}}\sin(\omega
t)$ and $V_B(t)=V_{B}^{\texttt{dc}}+V_{B}^{\texttt{ac}}\sin(\omega
t)$. However, the analytical results are valid for any contact
transparencies and driving potentials giving a proper turnstile
operation. The two quantum point contacts form a double barrier (DB)
with a set of resonant levels in between. The energies of the resonant
levels are taken to be time independent, controlled by the potential
$V_g$ applied to a top gate (see Fig.~\ref{device}).  The top-gate has
a large capacitance making charging effects negligible. \cite{Feve}
 
In this paper we focus on the energy distribution of the electrons
emitted towards terminal $2$, while the properties of the emitted holes could instead be investigated by
\textit{e.g.} changing the sign of the bias
$V$ at terminals $1$ and $3$. For the spectroscopy device we follow the edge
channel spectroscopy experiment in Ref. \onlinecite{pierre} and
consider a quantum dot weakly coupled to the output edge channel
leading to terminal $2$ (dashed box in Fig.~\ref{device}).  The
quantum dot has only one active level at energy $E_{spec}$, controlled
by a top gate voltage $V_{spec}$.  Electrons emitted by the turnstile
can tunnel through the quantum dot to an edge channel fed from a
grounded reservoir. The quantum dot acts as an energy filter and the
energy distribution of the emitted particles can be extracted from the
dc-component of the current at lead $M$. We point out that the
distance between the turnstile and the spectroscopy dot along the edge
is smaller than the energy relaxation length.\cite{pierrerelax, pierrerelax2}

\section{Floquet scattering approach}\label{floquetdynscattsec}
The energy distribution of the emitted particles is calculated within
the Floquet scattering approach. \cite{buttlong,buttpump,butttime} We first
focus on the energy distribution $\bar{f}_{out}(E)$ of the electrons
emitted from the turnstile propagating towards the spectroscopy
device, at a point denoted with $X$ in Fig.~\ref{device}. The relevant
scattering matrices are the Floquet transmission matrices from lead
$1$ to $X$, $\tilde{t}_{X1}(E)$, and from $4$ to $X$,
$\tilde{t}_{X4}(E)$. The element $t_{X\beta}(E_m,E_n)$ of the matrix
$\tilde t_{X\beta}$ is the amplitude for an electron incoming at
energy $E_n=E+n\hbar\omega$ from terminal $\beta=1,4$ to be emitted at
energy $E_m$ at $X$, picking up $m-n$ Floquet quanta $\hbar \omega$
when scattering at the time-dependent potentials. We have the matrices
\begin{eqnarray}
\tilde{t}_{X1}(E)&=&\tilde{t}_B\tilde{P}(E)[1-\tilde{r}_A\tilde{P}(E)\tilde{r}_B\tilde{P}(E)]^{-1}\tilde{t}_A\nonumber \\
\tilde{t}_{X4}(E)&=&\tilde{r}_B+\tilde{t}_B[1-\tilde{P}(E)\tilde{r}_A\tilde{P}(E)\tilde{r}_B]^{-1}\nonumber \\
&\times&\tilde{P}(E)\tilde{r}_A\tilde{P}(E)\tilde{t}_B.
\label{transmatrix}
\end{eqnarray}
The matrix $\tilde{P}(E)$ is diagonal with elements $P(E_m,E_m)=$
exp$[i\phi(E_m)]$. The phase $\phi(E_m)=\phi_0+\pi E_m/\Delta$ is
acquired when the particle, at energy $E_m$, propagates a length $L$
inside the DB, along the edge from $A$ to $B$ (or $B$ to $A$) at
drift velocity $v_D$. Here $\Delta=\pi\hbar v_D /L$ is the resonant
level spacing in the DB and $\phi_0$ is a constant phase, controlled
by $V_g$. The Floquet matrices $\tilde{t}_{A}, \tilde{r}_A$,
describing the scattering properties of quantum point contact $A$, are
taken energy independent on the scale $\mbox{max}\{kT, eV,
N_{max}\hbar\omega\}$, with $T$ the temperature and $N_{max}$ the
total number of contributing sidebands. The matrix elements $t_{A,nm}$
of $\tilde t_A$ are then given by the Fourier transform of the
time-dependent scattering amplitude $t_A(t)=i\sqrt{T_A(t)}$, \textit{i.e.}
\begin{equation}
t_{A,nm}=\frac{1}{\mathcal{T}}\int_0^{\mathcal{T}}e^{i(n-m)\omega t}t_A(t) dt,
\label{flelements}
\end{equation}
and similarly for $r_{A}(t)=\sqrt{1-T_{A}(t)}$. The matrices
$\tilde{t}_{B}, \tilde{r}_B$ describing the scattering properties of
$B$ are obtained in the same way. 

The distribution function $\bar{f}_{out}(E)$ is given by
\cite{buttpump} the quantum statistical average of the occupation
number of the outgoing edge at $X$. It can be written
\begin{equation}
\bar{f}_{out}(E)=\sum_{n}[T_{X1}^{n}(E)f_V(E_n)+T_{X4}^{n}(E)f_0(E_n)] 
 \label{fout}
\end{equation}
where $T_{X\beta}^{n}(E)=|t_{X\beta}(E,E_n)|^2$ and $f_V(E)$ and
$f_0(E)$ are the Fermi distribution functions of the biased and grounded
reservoirs respectively.

The experimental quantity of main interest is $I_{M}$, the direct, or
time averaged, part of the current flowing into terminal $M$ of the
spectroscopy device (see Fig.~\ref{device}). The spectroscopic quantum
dot is weakly coupled to the edges via two quantum point contacts $S$
and $D$ with transparencies $T_S, T_D\ll1$.  Since there is only one
active spin polarized level in the dot, Coulomb effects are
unimportant and the transport through the dot can be described as
transmission through a Breit-Wigner resonance. A calculation, along
the same line as for $\bar{f}_{out}(E)$, of the direct current flowing
into terminal $M$ then gives
\begin{equation}
 I_{M}=\frac{e}{h}\int T_{QD}(E)f_{out}(E)\mbox{ }dE 
\label{currM}
\end{equation}
where
$T_{QD}(E)=\Gamma_S\Gamma_D/[([\Gamma_S+\Gamma_D]/2)^2+(E-E_{spec})^2]$
with $\Gamma_{S/D}=T_{S/D}\Delta_{spec}/2\pi$ and $\Delta_{spec}$ the
level spacing of the dot.  The distribution function
$f_{out}(E)=\bar{f}_{out}(E)-f_0(E)$ is the difference between the
distribution function of the electron emitted by the turnstile
$\bar{f}_{out}(E)$ and the distribution function $f_0(E)$ of a
grounded reservoir. Importantly, to have a good resolution of the
energy distribution, the width of the quantum dot resonance must be
smaller than the energy scale $\delta E$ on which $f_{out}(E)$
changes, $\Gamma_S+\Gamma_D\ll\delta E$. In this limit we can
effectively take $T_{QD}(E) \propto \delta(E-E_{spec})$ and
Eq. (\ref{currM}) turns into
\begin{equation}
 I_{M}=\frac{e}{\hbar}\frac{\Gamma_S\Gamma_D}{\Gamma_S+\Gamma_D}f_{out}(E_{spec}). 
\label{currMdelta}
\end{equation}
Eq. (\ref{currMdelta}) shows explicitly that the direct current
flowing to lead $M$ is proportional to the energy distribution
function $f_{out}$ at energy $E_{spec}$. The full energy dependence of
the distribution can thus be reconstructed by continuously shifting
$E_{spec}$, achieved by tuning $V_{spec}$. \cite{pierre}

As will be clear from the discussion below, it is physically motivated
to part the energy distribution function $f_{out}(E)$ into two
contributions: one due to the applied bias
$f_{out}^{\texttt{bias}}(E)$,
\begin{equation}
f_{out}^{\texttt{bias}}(E)=\sum_{n}T_{X1}^{n}(E)[f_V(E_n)-f_0(E_n)]
 \label{foutbias}
\end{equation}
and one coming from the pumping effect $f_{out}^{\texttt{pump}}(E)$,
\begin{equation}
f_{out}^{\texttt{pump}}(E)=\sum_{n}[T_{X1}^{n}(E)+T_{X4}^{n}(E)][f_0(E_n)-f_0(E)]
 \label{foutpump}
\end{equation}
similar to the current partition in Ref. \onlinecite{Battista}.

\section{Frequency regimes}\label{freqregsec}
\begin{figure*}
\centerline{\psfig{figure=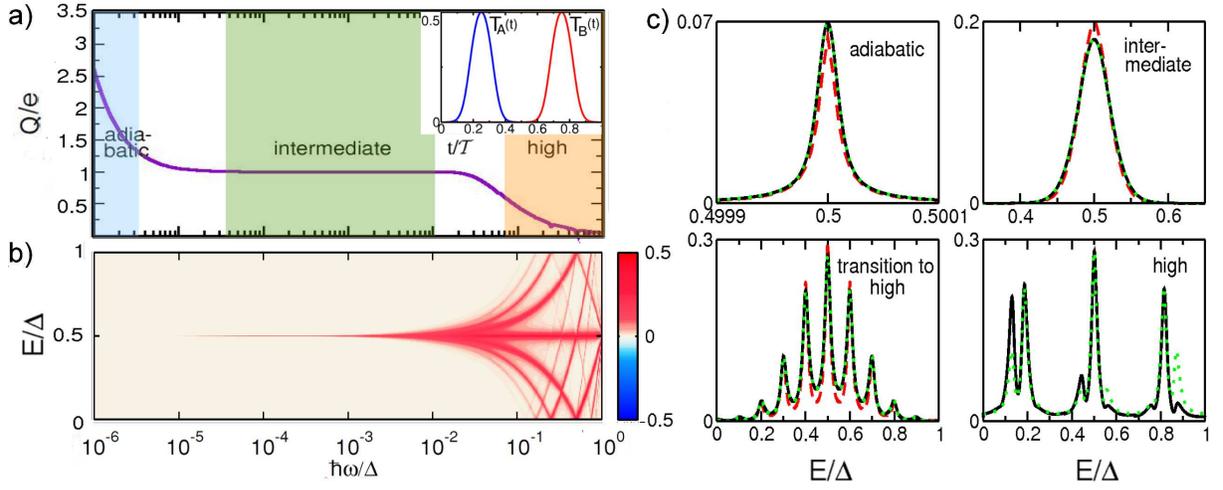,width=16cm}}
\caption{a) Charge $Q$ transferred by the turnstile per period, as a
  function of frequency with the three different regimes highlighted. Inset: Transparencies $T_A(t)$ and $T_B(t)$
  for the driving scheme used in the numerical calculations throughout
  the paper. b) Energy distribution $f_{out}(E)$ as a function of
  pumping frequency and energy.  c) Energy distribution $f_{out}(E)$ at
  $\hbar\omega=10^{-6}\Delta$ (adiabatic frequency), $\hbar\omega=10^{-2}\Delta$
  (intermediate frequency), $\hbar\omega=10^{-1}\Delta$ (transition to high frequency) and
  $\hbar\omega=10^{-0.5}\Delta$ (high frequency) with (solid black line) and
  without (dotted green line) pumping contribution. The red dashed lines 
show the analytical expressions obtained in the adiabatic [Eq. (\ref{adiabiasfoutwt})], intermediate [Eq. (\ref{foutistant})]
 and transition to high frequency regime [Eq. (\ref{fringes})].
 In all the plots $kT \ll \Delta$.}
\label{foutfig}
\end{figure*}
In the remaining part of the paper we focus on the distribution
function $f_{out}(E)$ and its parts $f_{out}^{\texttt{bias}}(E)$ and
$f_{out}^{\texttt{pump}}(E)$ in different frequency regimes. Based on
the findings in Ref. \onlinecite{Battista} we consider three
qualitatively different regimes with adiabatic, intermediate and high pumping
frequencies highlighted in Fig.~\ref{foutfig}~a). We will for simplicity focus our investigation on the
case with the applied voltage $eV=\Delta$ and one DB-resonance within
the bias window at energy $\epsilon_d=\Delta/2$. However the main
results of the paper rely only on the fact that there is a single
DB-level well inside the bias window.

\subsection{Adiabatic frequency regime}
In the adiabatic regime the pumping period $\mathcal{T}$ is much
longer than the time the particles spend inside the DB-region and many
particles flow through the turnstile during one period.
\cite{Battista} Formally
$N_{max}\hbar\omega\ll\Delta\mbox{min}[T_A(t)+T_B(t)]$, \textit{i.e.}
the total scattering matrix is energy independent on the scale
$N_{max}\hbar\omega$.  The numerically calculated energy distribution
function $f_{out}(E)$ is plotted as a function of frequency in
Fig.~\ref{foutfig}~b). We see that in the adiabatic regime the distribution
is sharply peaked around the DB-region resonance energy.
 
To obtain a quantitative estimate of the shape of the distribution
peak we note that to lowest order in frequency the Floquet scattering
matrix elements defined in Eq. (\ref{transmatrix}) are given by the
Fourier coefficients of the frozen scattering amplitude
\cite{buttpump} $t^{0}_{X\beta}(E,t)$ as
\begin{equation}
t_{X\beta}(E_n,E_m)=\frac{1}{\mathcal{T}}\int^{\mathcal{T}}_{0}e^{i(n-m)\omega t} t^{0}_{X\beta}(E,t)\mbox{ }dt.
 \label{adiaFloq}
\end{equation}
In our case the relevant frozen amplitudes are
\begin{eqnarray}
t_{X1}^{0}(E,t)&=&\frac{t_B(t)t_A(t)e^{i\phi(E)}}{1-r_A(t)r_B(t)e^{i2\phi(E)}}\nonumber \\
t_{X4}^{0}(E,t)&=&r_B(t)+\frac{r_A(t)t_B^2(t)e^{i2\phi(E)}}{1-r_A(t)r_B(t)e^{i2\phi(E)}}.
\label{frozmatrices}
\end{eqnarray}
Substituting Eq. (\ref{adiaFloq}) into
Eqs. (\ref{foutbias}-\ref{foutpump}) and, taking
$f_V(E_n)-f_0(E_n)\simeq f_V(E)-f_0(E)$, we find that the bias
contribution to the energy distribution is
\begin{equation}
f_{out}^{\texttt{bias},ad}(E)=\frac{1}{\mathcal{T}}\int_{0}^{\mathcal{T}} |t_{X1}^{0}(E,t)|^2[f_V(E)-f_0(E)]\mbox{ }dt.
\label{adiabiasfout}
\end{equation}
The pump component is found to be a factor $\omega/\Delta\ll1$ smaller
than the bias contribution $f_{out}^{\texttt{bias},ad}(E)$ and
$f_{out}^{\texttt{pump},ad}(E)$ is thus completely negligible; in
Fig.~\ref{foutfig}~c) $f_{out}(E)$ and $f_{out}^{\texttt{bias}}(E)$ fully
overlap.

It is known\cite{Battista} that the current flows
through the DB-region at times when the product $T_{A}(t)T_{B}(t)$ is
maximal, \textit{i.e.} around $t=n\mathcal{T}/2$. For those times
$T_{A}(t),T_{B}(t)\ll1$ and the electrons emitted by the turnstile
will be distributed in energy according to the time average of a
Breit-Wigner resonance at $\epsilon_d$, with time-dependent width
$\sim [T_{A}(t)+T_{B}(t)]\Delta$. Eq. (\ref{adiabiasfout}) then
simplifies to
\begin{equation}
f_{out}^{ad}(E)\simeq\frac{1}{\mathcal{T}}\int_0^{\mathcal{T}} \frac{T_A(t)T_B(t)}{\big(\frac{T_A(t)+T_B(t)}{2}\big)^2+\big(\frac{2\pi(E-\epsilon_d)}{\Delta}\big)^2}\mbox{ }dt
\label{adiabiasfoutwt}
\end{equation}
where we used that $f_V(E)-f_0(E)\simeq 1$ for the energies of
interest.  From the plot in Fig.~\ref{foutfig}~c) we see that
$f_{out}^{ad}(E)$ is in good agreement with the full numerics (the
small-shape discrepancy is due to the frequency not being in the
deep adiabatic regime).

\subsection{Intermediate frequency regime}\label{intermsubsec}
In the adiabatic regime, for increasing pumping frequency the number
of particles transversing the DB-region during one period decreases as
$1/\omega$. After a rapid transition into the non-adiabatic regime,
there is\cite{Battista} a wide pumping frequency interval $\Delta
\mbox{min}[T_A(t)+T_B(t)]/\hbar\ll\omega \ll
\omega_A^{\texttt{max}},\omega_B^{\texttt{max}}$ with $\hbar
\omega_{A/B}^{\texttt{max}}=\Delta \mbox{min}\{1,\int_0^{{\mathcal T}}
(dt/{\mathcal T})\ln[1/R_{A/B}(t)]\}$, in which the turnstile works
optimally and only one particle per period is pumped through the
turnstile.  As is clear from Fig.~\ref{foutfig}~b) and c)the energy distribution
of the electrons emitted in the optimal regime is still centered around
$\epsilon_d$, similar to the adiabatic regime, but it broadens and
changes shape.

Again, to obtain a quantitative expression for the distribution
function we first note that the pumping contribution
$f_{out}^{\texttt{pump}}(E)$ is still a factor $\omega/\Delta\ll1$
smaller than $f_{out}^{\texttt{bias}}(E)$.  We can thus write
\begin{equation}
f_{out}(E)\simeq f_{out}^{\texttt{bias}}(E)\simeq\sum_{n}T_{X1}^{n}(E)[f_V(E)-f_0(E)].
 \label{foutbiasint}
\end{equation}
For the energies of interest, around $\epsilon_d$,
$f_V(E)-f_0(E)\simeq 1$. To be able to treat arbitrary, non-adiabatic frequencies we
introduce the dynamical scattering
amplitude $t_{X1}(E,t)$, defined as \cite{butttime}
\begin{equation}
  t_{X1}(E,t)=\sum_n e^{in\omega t}t_{X1}(E,E_n),
\label{dynscatdef}
\end{equation}
for an electron injected from terminal $1$ at time $t$ to be emitted
to $X$ with energy $E$. We can then write Eq. (\ref{foutbiasint}) as
\begin{equation}
 f_{out}(E)=\frac{1}{\mathcal{T}}\int_{0}^{\mathcal{T}} |t_{X1}(E,t)|^2 dt.
 \label{foutbiasdyn}
\end{equation}
From Eqs. (\ref{transmatrix}) and (\ref{dynscatdef}) we obtain
(similar to Ref. [\onlinecite{butttime}])
\begin{equation}
 t_{X1}(E,t)=t_A(t)\sum_{q=0}^{\infty}e^{i(2q+1)\phi(E)}
L_{q}(t)t_B(t+[2q+1]\tau), 
\label{dynsout}
\end{equation}
where $\tau=L/v_D$ and $L_q(t)=\prod_{p=1}^q
r_{A}(t+2p\tau)r_B(t+[2p-1]\tau)$ for $q\geq 1$ and $1$ for $q=0$.  In
the intermediate regime the time of flight $\tau$ through the
DB-region is much smaller than the pumping period,
$\tau\ll\mathcal{T}$.  We can then go from a discrete to a continuous
description in time and write $t_{X1}(E,t)$ in Eq. (\ref{dynsout}) as
\begin{eqnarray}
 t_{X1}(E,t)&\simeq& -t_A(t)\frac{1}{2\tau}\int_0^{\infty}e^{i\frac{(E-\epsilon_d)}{\hbar}t'}t_B(t+t')\nonumber \\
&\times&e^{\frac{1}{2\tau}\int_0^{t'}\ln[r_A(t+t'')r_B(t+t'')]dt''}\mbox{ }dt'.
\label{dynsout2}
\end{eqnarray}
In the optimal turnstile regime we can neglect \cite{Battista} the times when both
contacts $A$ and $B$ are simultaneously open and put $t_B(t)=0$ for
$0<t<\mathcal{T}/2$ and $t_A(t)=0$ for
$\mathcal{T}/2<t<\mathcal{T}$. Moreover, the current flows in and out
of the DB-region when the respective quantum points contacts are
starting to open, \textit{i.e.} $T_A(t),T_B(t)\ll1$, and we can expand
the logarithm in Eq. (\ref{dynsout2}) to first order in
$T_{A}(t), T_{B}(t)$. We can then write the quantity of interest
 \begin{equation}
   |t_{X1}(E,t)|^2=C_A(t)|c_B(E)|^2,
\label{dynsout4}
\end{equation}
where we can identify
\begin{equation}
 C_A(t)=T_A(t)e^{-\frac{1}{2\tau}\int_t^{\frac{\mathcal{T}}{2}}T_A(\bar{t})\mbox{ }d\bar{t}}
\end{equation}
as the probability that an electron is injected at $t$ and thereafter
stays inside the DB-region until $\mathcal{T}/2$. The energy dependent
function
\begin{equation}
c_B(E)=\frac{1}{2\tau}\int_{\frac{\mathcal{T}}{2}}^{\infty}e^{i\frac{(E-\epsilon_d)}{\hbar}t'}t_B(t')e^{-\frac{1}{4\tau}\int_{\frac{\mathcal{T}}{2}}^{t'}T_B(t'')dt''}dt'
\label{cB}
\end{equation}
depends on the scattering properties of contact $B$ only. The distribution function, Eq. (\ref{foutbiasdyn}), is then
\begin{equation}
  f_{out}(E)= \frac{\hbar \omega}{\Delta}|c_B(E)|^2
\label{foutistant}
\end{equation}
since the time integral
\begin{equation}
\frac{1}{2\tau}\int_0^{\frac{\mathcal{T}}{2}} C_A(t) dt=1-e^{-\frac{1}{2\tau}\int_0^{\frac{\mathcal{T}}{2}}T_A(\bar{t}) d\bar{t}}\simeq 1
\end{equation}
gives \cite{Battista} the probability that the DB-region is charged at $\mathcal{T}/2$, unity in the optimal pumping regime. 

The expression for $f_{out}(E)$ in Eq. (\ref{foutistant}), together with
Eq. (\ref{cB}), allows for a straightforward evaluation of the
distribution function of the emitted electrons for any ideal turnstile
driving scheme, once the time dependent transparency $T_B(t)$ is
known. As is clear from Fig.~\ref{foutfig}~c), the expression in
Eq. (\ref{foutistant}) gives very good agreement with the full
numerics.

\subsubsection*{Floquet fringes}

At frequencies $\omega_{A/B}^{\texttt{max}}<\omega<\Delta/\hbar$ the
transferred charge starts to decrease, $Q<1$ [see Fig.~\ref{foutfig}~a)],
a consequence of incomplete charging and discharging of the DB-region
during the pumping cycle.  In this frequency regime
$f_{out}^{\texttt{bias}}(E)$ still determines the spectral
distribution of the emitted charge, the pumping contribution
$f_{out}^{\texttt{pump}}(E)$ can be neglected as the numerics show in
Fig.~\ref{foutfig}~c).

From Fig.~\ref{foutfig}~b) we note that the resonance at $\epsilon_d$
starts to split up, \textit{i.e.} $f_{out}(E)$ develops a set of
fringes at energies $\pm n\hbar\omega$ around $\epsilon_d$, with
$n=0,1,2,3...$.  The fringes are a manifestation of the integer number
of Floquet quanta which the electrons gain or loose when transmitting
through the DB-region. Such manifestations of the interactions of
transport electrons with an applied, time-dependent field has been
intensively investigated in various forms in mesoscopic conductors,
see \textit{e.g.} Refs. \onlinecite{bruder,patkouvenI,patkouvenII,pedersen} for
early works and Ref. \onlinecite{Platero} for a review. Typically the
effect of the time-dependent field was investigated via transport
quantities such as average current, differential conductance or
noise. Here we focus on the manifestation of the time-periodic field,
\textit{i.e.} the Floquet fringes, directly in the distribution function.
\begin{figure}[h]
\centerline{\psfig{figure=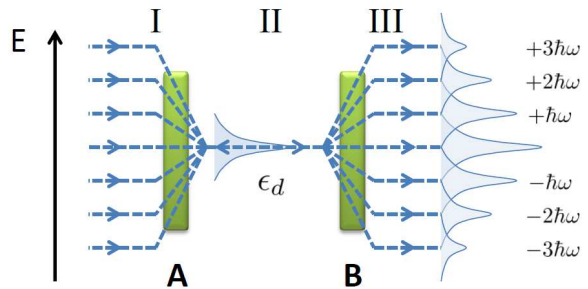,width=8cm}}
\caption{Schematic of the resonant paths in energy-position space, with three regions I, II and III. In I electrons incident from terminal 1 pick up or lose a given number of Floquet quanta at contact $A$, to hit the resonance energy $\epsilon_d$. In II, the DB-region between $A$ and B, the electrons scatter elastically back and forth between the contacts. In III the electrons acquire $0,\pm 1,\pm 2...$ quanta when transmitting out through B.}
\label{fringefig}
\end{figure} 

To obtain a quantitative description of the Floquet fringes we note
that for well separated fringes, only electrons which scatter
resonantly through the DB contribute significantly to
$f_{out}(E)$. The resonant paths in energy space are shown in
Fig.~\ref{fringefig}. One can divide the resonant process into three
subsequent parts: I) only electrons incident with an energy around
$\epsilon_d\pm p\hbar \omega$, \textit{i.e.} such that they can hit the resonance
at $\epsilon_d$ by losing or gaining $p=0,1,2,..$ Floquet quanta
$\hbar \omega$ at contact $A$, can enter the DB-region. II) inside the
DB-region electrons scatter elastically back and forth between $A$ and
B, \textit{i.e.} without acquiring any quanta. III) electrons emitted out
through contact $B$ pick up or lose $n$ quanta and thereby contribute to
the fringes at $\epsilon_d \pm n\hbar \omega$. Summing up all resonant
paths we have from Eq. (\ref{foutbias}) and Eqs. (\ref{transmatrix}),
(\ref{flelements}) that 
\begin{equation}
f_{out}(E)=\frac{\Delta}{2\pi}\sum_n \frac{\bar \Gamma_A |t_{B,0n}|^2}{([\bar\Gamma_A+\bar \Gamma_B]/2)^2+(E_n-\epsilon_d)^2}
\label{fringes}
\end{equation}
where we use that $f_V(E_n)-f_0(E_n)\simeq 1$ and introduced $\bar
\Gamma_{A/B}=(1/{\mathcal T})\int_0^{\mathcal T} dt \Gamma_{A/B}(t)$,
the time average of the tunneling rate $\Gamma_{A/B}(t)=T_{A/B}(t)\Delta/2\pi$. The expression in
Eq. (\ref{fringes}) gives good agreement with the full
numerics as shown in Fig.~\ref{foutfig}~c). From Eq. (\ref{fringes}) we also see
that fringes are given by a set of Lorentzians centered around
$\epsilon_d+n\hbar \omega$, with a width $\bar\Gamma_A+\bar \Gamma_B$.
The peak height of the fringes are proportional to $|t_{B,0n}|^2$, the
modulus square of the Fourier components of the transmission amplitude
through contact B. We point out that further numerical investigations
(not presented) show that the occurrence of fringes of $f_{out}(E)$
given by Eq. (\ref{fringes}) is a generic feature for a turnstile with
a single active DB-level and hence not due to the specific parameters
used in Fig.~\ref{foutfig}~b).

\subsection{High frequency regime}
\begin{figure}
\centerline{\psfig{figure=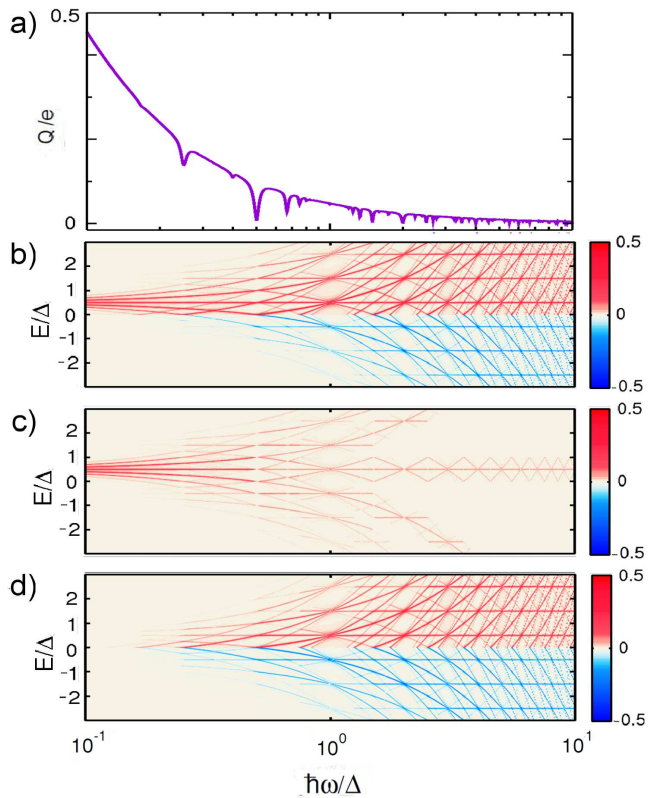,width=8.5cm}}
\caption{a) High frequency regime of the transferred charge $Q$ per period, displaying dips at
  frequencies given by Eq. (\ref{biasfringes}). b)-d) Energy distributions as functions of energy 
  and drive frequency.  The total distribution $f_{out}(E)$, [in b)], the bias
  contribution $f_{out}^{\texttt{bias}}$, [in c)] and the pumping contribution
  $f_{out}^{\texttt{pump}}(E)$ [in d)] are shown. In all the plots $T_A(t),T_B(t)$ as in
  Fig.~\ref{foutfig} and $kT \ll \Delta$. }
  \label{dips}
\end{figure}
In the high frequency regime the Floquet quantum $\hbar\omega$ becomes
comparable to the DB-level spacing $\Delta$. As a result the Floquet
fringes from electrons scattering through DB-resonances at energies
$\epsilon_d + m\Delta$, $m=\pm 1,\pm 2...$ start to contribute to
$f_{out}(E)$. As shown in Fig.~\ref{dips}, this leads to a dense
pattern of fringes moving up and down in energy with increasing
frequency $\omega>\Delta/\hbar$. In contrast to the adiabatic and
intermediate frequency regimes, in the high frequency regime the
pumping contribution $f_{out}^{\texttt{pump}}(E)$ and the bias
contribution $f_{out}^{\texttt{bias}}(E)$ are comparable [see Fig.~\ref{foutfig}~c)].

It is helpful for the physical understanding to discuss the properties
of the two contributions separately. Starting with the bias
contribution, we first note that $f_{out}^{\texttt{bias}}(E)$ is
manifestly positive [see Eq. (\ref{foutbias})], describing Floquet
scattering of electrons injected from terminal 1, in the bias window.
Moreover, with the resonance $\epsilon_d=\Delta/2$ in the middle of
the bias window, the symmetry
$f_{out}^{\texttt{bias}}(\epsilon_d+E)=f_{out}^{\texttt{bias}}(\epsilon_d-E)$
follows from Eq. (\ref{foutbias}) and is directly visible in
Fig.~\ref{dips}~c). The energy of the fringes in
$f_{out}^{\texttt{bias}}$ can be found by extending the reasoning
above: an incident electron which scatter through a DB-resonance at
$\epsilon_d + m\Delta$ and emit an additional $n=0,\pm 1, \pm 2,...$
quanta when transmitting out through contact $B$ contributes to a fringe
at an energy $E=\epsilon_{nm}$, given by
\begin{equation}
\epsilon_{nm}=\epsilon_d+m\Delta+n\hbar\omega.
\label{biasfringes}
\end{equation}
Since the incident energy of the electron is restricted to the bias
window, the bias component $f_{out}^{\texttt{bias}}$ will only show
fringes at energies $\epsilon_{nm}$ fulfilling the additional
requirement $\hbar \omega p \leq \epsilon_{nm} \leq eV+\hbar \omega
p$, with $p=0, \pm1, \pm2...$ For $\hbar \omega > eV$ this leads to
bands of fringes, $eV$ wide and separated by $\hbar \omega$, as is
clearly seen in Fig.~\ref{dips}~c).

Turning to the pump contribution, $f_{out}^{\texttt{pump}}(E)$
describes the creation of electron-hole pairs out of the Fermi sea,
due to the time dependent potentials at the quantum point contacts $A$
and $B$. The pump contribution has the symmetry
$f_{out}^{\texttt{pump}}(E)=-f_{out}^{\texttt{pump}}(-E)$, also
visible in Fig.~\ref{dips}~d). The origin of the fringes in
$f_{out}^{\texttt{pump}}(E)$ is the same as for the bias part,
however, since all electrons below Fermi energy in principle can
contribute, in contrast to the bias part the fringes appear at all
energies $\epsilon_{nm}$ given by Eq. (\ref{biasfringes}).

Common for the fringes in $f_{out}^{\texttt{bias}}(E)$ and
$f_{out}^{\texttt{pump}}(E)$ is that they typically are described as a
set of superimposed Lorentzians of width $\bar \Gamma_A+\bar \Gamma_B$ in
energy, just as described above for the intermediate frequency regime
[see Eq. (\ref{fringes})]. However, the height of the peaks depend in
a more complicated way on through which resonances the particles have
scattered as well as on the available energies for the injected
electrons, giving a more complex peak structure. This is clear from
the high frequency regime panel in Fig.~\ref{foutfig}~c), where
$f_{out}^{\texttt{bias}}(E)$ as well as the total distribution
$f_{out}(E)$, the experimentally accessible quantity, are shown.

\subsubsection{Relation to transferred charge}
\begin{figure}
\centerline{\psfig{figure=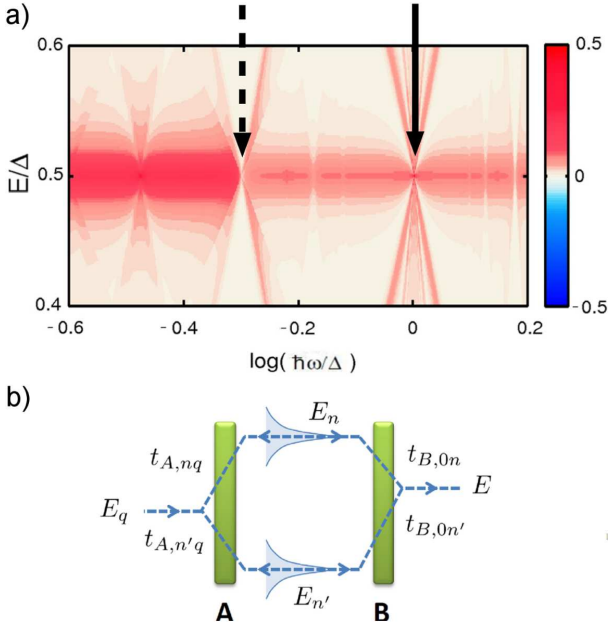,width=8cm}}
\caption{a) Close-up of distribution function $f_{out}^{\texttt{bias}}(E)$ around 
frequencies $\hbar\omega\sim\Delta$, for energies around resonance $\epsilon_d=\Delta/2$. 
Strong suppression due to destructive interference is clear for $\hbar\omega=\Delta/2$ (dashed arrow)
while no suppression occurs for $\hbar\omega=\Delta$ (full line arrow). b) 
Schematic of two energy-position paths, going through two different resonances.}
  \label{resfig}
\end{figure}

It is interesting to relate the fringe properties of $f_{out}(E)$ and
its components to the transferred charge $Q$ per
cycle, discussed in Ref. \onlinecite{Battista} and plotted as a
function of frequency for reference in Fig.~\ref{dips}~a). In the high
frequency regime, in addition to a slow, $\sim 1/\omega$, overall
decrease with increasing frequency, the charge $Q$ displays sharp dips
at certain frequencies. In Ref. \onlinecite{Battista} these dips were
explained by appealing to semi-classical electron paths through the
turnstile leading to zero or small charge transfer. Here we first note
that the charge $Q$ is determined by $f_{out}^{\texttt{bias}}(E)$
only, since $Q=(e/h)\mathcal{T}\int f_{out}(E) dE$ and $f_{out}^{\texttt{pump}}(E)$
is anti-symmetric in energy around $E=0$. By a direct comparison of
the fringe structure of $f_{out}^{\texttt{bias}}(E)$ in
Fig.~\ref{dips}~c) with the dips of $Q$ in Fig.~\ref{dips}~a), we note
that all dips occur for frequencies where different fringes cross, \textit{i.e.}
[from Eq. (\ref{biasfringes})] when $\epsilon_{nm}=\epsilon_{n'm'}$
with $n\neq n', m\neq m'$ giving the frequencies
\begin{equation}
\hbar \omega=\frac{m-m'}{n'-n}\Delta.
\label{dippos}
\end{equation}
However, not all fringe crossings correspond to charge dips,
\textit{e.g.} while for $\hbar\omega=\Delta/2$ the dip in $Q$ is large, for
$\hbar\omega=\Delta$ there is no dip at all. The fringe structure around
these two frequencies is illustrated in detail in Fig.~\ref{resfig}~a).

To understand both the position and magnitude of the dips we note that
crossing fringes correspond to a situation when an injected electron
can take two (or more) different resonant paths in energy space and be
emitted at the same energy $E$, see illustration in
Fig.~\ref{resfig}~b). As a consequence the two paths interfere,
constructively or destructively, depending on the relative amplitudes
for the two paths. To illustrate this which-energy-path interference \cite{buttpump}
we consider two different paths where an electron injected at energy
$E_q$ scatter through a resonance at energy $E_n=\epsilon_d+m\Delta$
or $E_{n'}=\epsilon_d+m'\Delta$, thereafter loses/gain $n$ or $n'$ quanta
respectively and is emitted at energy $E$. The contribution to
$f_{out}^{\texttt{bias}}(E)$ for this process is then, similar to
Eq. (\ref{fringes}), exactly at resonance
\begin{equation}
4\frac{|t_{A,nq}t_{B,0n}e^{i\pi m \hbar \omega/\Delta}+t_{A,n'q}t_{B,0n'}e^{i\pi m' \hbar \omega/\Delta}|^2}{(\bar\Gamma_A+\bar \Gamma_B)^2}.
\label{fringeint}
\end{equation}
For the symmetric turnstile with $\pi$-out of phase driving considered
here we have $T_A(t)=T_B(t+{\mathcal T}/2)$ and consequently
$t_{A,mn}=t_{B,mn}e^{i(m-n)\pi}$. The ratio of the interference, or coherent part and the incoherent part is then given by 
\begin{equation}
\frac{2t_{B,nq}t_{B,0n}t_{B,n'q}t_{B,0n'}\cos([m-m']\pi[1+\hbar\omega/\Delta])}{|t_{B,nq}t_{B,0n}|^2+|t_{B,n'q}t_{B,0n'}|^2}
\label{fringeint2}
\end{equation}
Noting that $t_{B,mn}$ is purely imaginary, this ratio gives
\textit{e.g.} $-1$ for all fringe crossings at $\hbar \omega=\Delta/2$,
complete destructive interference, while it is $1$, complete
constructive interference, for all fringe crossings at $\hbar
\omega=\Delta$ (where more than two fringes cross), see
Fig.~\ref{resfig}~a). For most crossings the ratio is somewhere in
between, due to different probabilities for the individual paths,
\textit{i.e.} $|t_{A,nq}t_{B,0n}|^2 \neq |t_{A,n'q}t_{B,n'0}|^2$. Since all fringe
crossings for a given frequency (at different energies) have the same
cosine-factor [see Eq. (\ref{fringeint2})], all crossings contribute with the same sign. Taken
together, the energy-path interference picture provides a quantum
mechanical explanation for both the origin and the magnitude of the
dips in the transferred charge $Q$, complementing and extending the
semiclassical explanation given in Ref. \onlinecite{Battista}.

\section{Wavefunction of emitted electrons, optimal turnstile
  regime.}\label{wfsec}
Complete information about the state emitted by the turnstile is
obtained from the full many-body wavefunction. The wavefunction is of
key importance when investigating the possibilities for quantum
information processing with electrons in the quantum Hall regime. Of
particular interest is the wavefunction for the electrons emitted in
the optimal regime, with exactly one electron transferred per cycle.
It would be desirable to derive the many-body wavefunction in the
optimal regime along the lines of Ref. \onlinecite{Levitov}, where it
was formally shown that the pump in the F\`{e}ve \textit{et
  al.}\cite{Feve} experiment under ideal conditions creates a single
electron (or hole) excitation on top of a filled Fermi sea. Here we
however take a simpler path, which we nevertheless argue gives the
same result in the optimal pumping regime.

During optimal operation the turnstile is completely charged and
subsequently discharged, once each period.
\begin{figure}[h]
\centerline{\psfig{figure=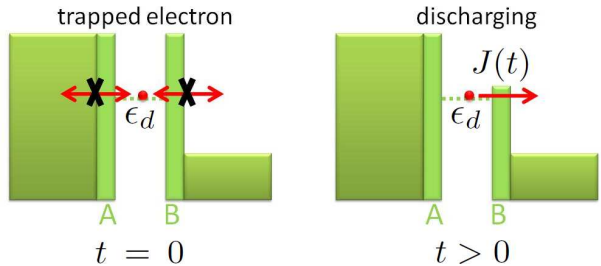,width=8cm}}
\caption{Sketch of the simplified model used to derive the single
  particle wavefunction in the ideal regime. At $t=0$ the electron is
  trapped between barriers, in a single level at energy
  $\epsilon_d=0$. For $t>0$ the electron can start to tunnel to the
  right lead. The time-dependent lead-level coupling is $J(t)$.}
\label{wfpic}
\end{figure}
In the second half of the pumping period, during the discharging, the
quantum point contact $A$ is closed and the electron trapped in the DB
can escape out through $B$ (as sketched in Fig.~\ref{wfpic}).  For
$\hbar\omega\ll\Delta$ the escape takes place during times when
$T_B(t)\ll1$, \textit{i.e.} soon after the opening of $B$ at
$t=\mathcal{T}/2$. Then only energies in a narrow interval $\sim
T_B(t)\Delta$ around $\epsilon_d=\Delta/2$ are of importance. This
allow us to neglect many-particle processes, related to excitations
out of the filled Fermi sea and consider a simplified single-particle
model for the wavefunction of the emitted state. We thus describe the
system during discharging with the Hamiltonian
\begin{equation}
 H=\epsilon_d|d\rangle\langle d|+\sum_E E|E\rangle\langle E|+\sum_E J(t)\big[|d\rangle\langle E|+|E\rangle\langle d|\big]
\label{hami}
\end{equation}
where $|d\rangle$ denotes the DB-level and $|E\rangle$ the chiral edge
state of energy $E$, outside the DB-region.  The tunnel coupling
$J(t)$ between the DB level and the edge is assumed to be energy
independent.  For notational convenience, in the model we count
energies away from the dot resonance [$\epsilon_d=0$ in
Eq. (\ref{hami})] and take the onset of the tunneling to occur at
$t=0$.  We substitute the ansatz
\begin{equation}
 |\psi(t)\rangle=c_d(t)|d\rangle+\sum_E c_E(t)|E\rangle
\label{ansatz}
\end{equation}
into the time dependent Schr\"{o}dinger equation $i\hbar
d|\psi(t)\rangle/dt=H |\psi(t)\rangle$ and get the system of equations
\begin{eqnarray}
i\hbar \dot{c}_d(t)&=&\sum_E J(t)c_E(t), \nonumber \\ 
i\hbar \dot{c}_E(t)&=& E c_E(t) +J(t)c_d(t),
\label{eqsystem}
\end{eqnarray}
subjected to the initial condition
\begin{equation}
c_d(t=0)=1,\hspace{1cm}c_E(t=0)=0.
\label{initialcond}
\end{equation}
Eqs. (\ref{eqsystem}) with the boundary conditions in
Eq. (\ref{initialcond}) can conveniently be solved by means of a
Laplace transformation (see Appendix \ref{appendixB} for details).
For a continuum of states outside the DB, we find
\begin{equation}
c_d(t)=e^{-\int_{0}^{t} \Gamma(\bar{t})\mbox{ }d\bar{t}},\hspace{1cm}\Gamma(t)=\frac{\pi\nu}{\hbar}J^2(t)
 \label{cdt}
\end{equation}
where $\nu$ is the density of states of the edge.  Using this
expression for $c_d(t)$ we can solve the remaining equations in
(\ref{eqsystem}) and find
\begin{equation}
c_E(t)=e^{\frac{-iEt}{\hbar}}\int_{0}^t e^{\frac{iEt'}{\hbar}}\frac{J(t')}{i\hbar}e^{-\int_{0}^{t'} \Gamma(\bar{t})\mbox{ }d\bar{t}} dt'.
 \label{cEt}
\end{equation}
At times much longer than the emission time the DB-region is
completely discharged, $c_d(t)\rightarrow0$, and the wavefunction
$|\psi(t)\rangle$ only describes the emitted electron in the edge, a
wavepacket
\begin{equation}
 |\psi(t)\rangle=\sqrt{\nu}\int c_E(t)|E\rangle dE.
\label{longtimewf}
\end{equation}
To employ this result for composite systems, with \textit{e.g.} two or more
turnstiles, it is desirable to construct the full many-body state
corresponding to $|\psi(t)\rangle$. Reintroducing the Fermi sea, in
second quantization we have
\begin{equation}
 |\psi\rangle=\int_0^{\infty} \tilde{c}_E a^{\dagger}_E|0\rangle dE
\label{wf}
\end{equation}
in the Heisenberg picture. Here $a^{\dagger}_E$ creates an electron in
the edge towards $X$ in Fig.~\ref{device} at energy $E$, $|0\rangle$
is the filled Fermi sea. The coefficient 
$\tilde{c}_E=\sqrt{\nu}/(i\hbar)\int_{0}^t e^{\frac{i(E-\epsilon_d)t'}{\hbar}}J(t')e^{-\int_{0}^{t'} \Gamma(\bar{t})\mbox{ }d\bar{t}} dt'$
reintroducing the turnstile level energy $\epsilon_d=\Delta/2$.
From the wavefunction in Eq. (\ref{wf}) the average occupation
number at energy $E$ is given by $\langle \hat n(E)\rangle\equiv
\langle a^{\dagger}_Ea_E\rangle=|\tilde{c}_E|^2$.

To connect the wavefunction result with the Floquet approach above we
first compare the probability for an electron to remain inside the DB
after opening contact B, $|c_d|^2$, with the result of
Ref. \onlinecite{Battista}. We find, shifting the onset of the tunneling $\mathcal{T}/2$ in time in $J(t)$, 
\begin{equation}
 J(t)=\frac{1}{2\pi}\sqrt{\frac{\Delta T_B(t)}{\nu}}, \hspace{0.5cm} t \in [0,{\mathcal T}/2].
\end{equation}
We can then compare the distribution function in the optimal regime in
Eq. (\ref{foutistant}) with the average occupation number from the
wavefunction, giving
\begin{equation}
  f_{out}(E)=\hbar \omega\langle\hat{n}_E\rangle
\label{foutistantwf}
\end{equation}
The factor $\hbar \omega$ in front of $\langle\hat{n}_E\rangle$ simply
reflects the fact that while the wavefunction $|\psi\rangle$ describes
a single electron emission, $f_{out}(E)$ describes the periodic
emission of single electrons, with a frequency $\omega$. The relation
in Eq. (\ref{foutistantwf}) provides evidence that the full manybody
wavefunction for a single discharging event is given by
Eq. (\ref{wf}). To obtain the wavefunction for several emitted
electrons, well separated in time, one acts upon $|0\rangle$ with a
product of wave packet operators $\int_0^{\infty} dE \tilde c_E
a_E^{\dagger}$ with time translated tunnel couplings $J(t)$ [or
equivalently $T_B(t)$], describing different emission times.

Importantly, the single particle wavefunction for the emitted electron
in Eq. (\ref{longtimewf}) is valid for arbitrary tunnel coupling
$J(t)$. However, we emphasize that special care must be taken when
making the connection to the manybody wavefunction in
Eq. (\ref{wf}). This is clearly illustrated by considering a steplike
onset at $t=0$, \textit{i.e.} $J(t)=J\theta(t)$. This gives an amplitude
\begin{equation}
 \tilde{c}_E=\sqrt{\frac{\hbar}{\pi}}\frac{\sqrt{\Gamma}}{i\hbar\Gamma+(E-\epsilon_d)}
\end{equation}
\textit{i.e.} a Lorentzian wavepacket centered around $\epsilon_d$. The problem
is that such a wavepacket is not well confined inside the bias window
$0 \leq E \leq \Delta$, the occupation decays as $\sim 1/E^2$ far away
from resonance. As a consequence there is a non-negligible probability
to find the electron inside the filled Fermi sea or in a higher lying
DB-level, incompatible with the assumptions for the optimal pumping
regime. This demonstrates that to make the connection between the
wavefunctions in Eqs. (\ref{longtimewf}) and (\ref{wf}), the
time-dependence of $J(t)$, and hence $T_B(t)$, has to be such that the
resulting single particle wavepacket has no spectral weight outside
the bias window.

\section{Imperfections and robustness}\label{imperfectionsec}
To assess the feasibility of our proposal it is important to
investigate possible imperfections or deviations from the model which
might become important in an experiment.  In our opinion, the most
important issue is the various effects of the capacitive coupling
between the different components in the system, \textit{e.g.} the
gates, the reservoirs and the electrons in the DB- region.

First and foremost, we have so far in the paper assumed that the
electrostatic potential of the DB is constant in time, due to a
dominating capacitive coupling to the metallic top gate kept at a
constant potential $V_g$ (see Fig.~\ref{device}). In an experiment,
this might not be the case and it is interesting to investigate the
effect of a capacitive coupling also to the gate at $A$ and $B$, with
applied time dependent potentials $V_A(t)$ and $V_B(t)$.  Second, a
capacitive coupling between the gates at $A$ and $B$ and the
electronic reservoirs $1$ to $4$ introduces a time dependent component
of the bias potential at the reservoirs. Such a time-dependent
potential can lead to a rectification current \cite{rectbutt} which
can obscure the physical phenomena under investigation. \cite{switkes,
  Brouwerrect}

Starting with the latter type of coupling, in our proposed turnstile,
the rectification effects are typically not important in the adiabatic
and in the intermediate frequency regimes. The reason for this is that
a small ac-potential at the reservoirs only leads to electron-hole
excitations around energies $0$ and $eV$ of the grounded and biased
reservoirs respectively. These energies are far away from the
resonance $\epsilon_d=\Delta/2$ where the net transport takes
place. Hence, similar to the pumping contribution
$f^{\texttt{pump}}_{out}(E)$, the rectification effects are negligible
for frequencies $\omega\ll\Delta/\hbar$.  At high frequencies
$\hbar\omega\sim\Delta/e$ rectification effects might become
important, their magnitude depends on the strength of the capacitive
coupling between the gates at $A$ and $B$ and the reservoirs. A
detailed investigation of these issues is however outside the scope of
this article.

\begin{figure}[h!]
\centerline{\psfig{figure=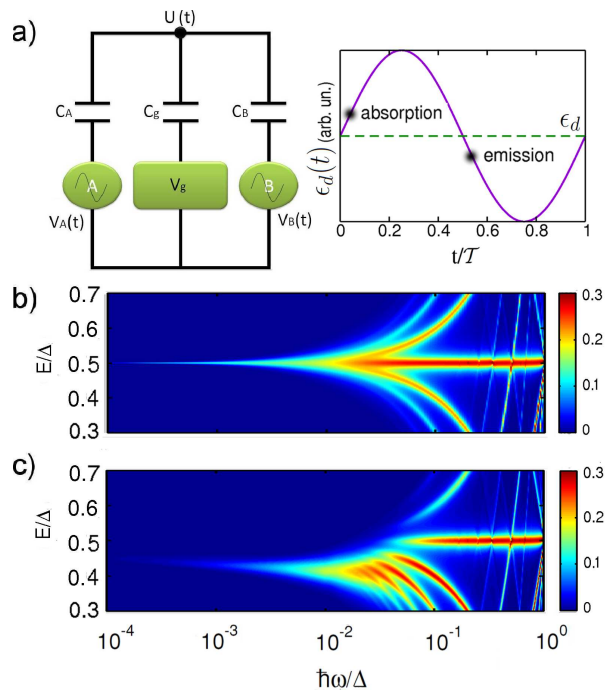,width=8cm}}
\caption{a) Left: Circuit representation of the turnstile DB-region
  capacitively coupled to quantum point contact gates $A$, $B$ subjected to
  time dependent bias $V_{A/B}(t)$ and a top gate kept at constant
  bias $V_g$. Right: sketch of the oscillation of the effective,
  instantaneous resonant level energy $\epsilon_d(t)$. Times for
  absorption and emission of the electron from the DB-region for typical system
  parameters are shown with black dots. b) Energy distribution $f_{out}(E)$ for asymmetry factor $e\alpha V^{\texttt{ac}}=0$
  to compare with c) $f_{out}(E)$ for $e\alpha V^{\texttt{ac}}=0.1\Delta$ for a selected interval of $E$ and $\omega$.}
\label{coupling}
\end{figure}

For the first type of coupling, inducing a time dependent potential in
the DB, the situation is a priori less clear and we therefore
investigate it in more detail. The isolated system, consisting of the
spatially constant DB-region capacitively coupled to the two quantum
point contact gates $A$ and $B$ and the DB-region top-gate, can be
represented in a simple circuit theory model as three capacitors of
capacitances $C_A,C_g,C_B$ put in parallel [see
Fig.~\ref{coupling}~a)]. Each of the capacitors is subjected to a
different voltage, $V_A(t), V_g, V_B(t)$ respectively, with $V_
{A/B}=V_{A/B}^{\texttt{dc}}\pm V_{A/B}^{\texttt{ac}}\sin(\omega t)$
discussed above.
The effective potential that the electron will experience in the
DB-region is $U(t)=U_0+\delta U(t)$. For a typical driving scheme
$V_A^{\texttt{ac}}=V_B^{\texttt{ac}}=V^{\texttt{ac}}$ we find
\begin{equation}
  \delta U(t)=\alpha V^{\texttt{ac}}\sin(\omega t), \hspace{0.5cm} \alpha=\frac{C_B-C_A}{C_A+C_g+C_B}.
\label{ut}
\end{equation}
The constant part $U_0$ (determining $\phi_0$ discussed above) is
determined by the top gate potential $V_g$.  Eq. (\ref{ut}) shows that
in case of a dominant coupling with the gate, $C_g\gg C_A,C_B$, the
asymmetry parameter $\alpha \ll 1$ and we can neglect $\delta U(t)$,
as was done in the previous sections. Furthermore due to the
$\pi-$shifted driving between $A$ and B, the induced time dependent
potential is proportional to the difference $C_B-C_A$. For a symmetric
capacitive coupling, $C_A=C_B$, we can thus also neglect $\delta
U(t)$.  To have an effect on $f_{out}(E)$ the capacitive $A,B$-gate
couplings thus have to be sizably asymmetric and of comparable
strength to the DB-top gate coupling.

To calculate the effect of $\delta U(t)$ we note that a time
dependent, spatially constant, potential in the DB can be formally
taken into account \cite{butttime} by modifying the Floquet matrix
$\tilde{P}(E)$ in Eq. (\ref{transmatrix}) to
\begin{eqnarray}
&&P(E_n,E_m)=\frac{1}{\mathcal{T}}\int^{\mathcal{T}}_{0}e^{i(n-m)\omega t} dt\nonumber \\
&\times& \exp\left(i\left[\phi_0+\frac{\pi E_n}{\Delta}-\frac{e}{\hbar}\int_t^{t+\tau} \delta U(t')dt'\right]\right)
 \label{nondiagP}
\end{eqnarray}
which is then no longer diagonal. With the modified expression for the
transmission matrices in Eq. (\ref{transmatrix}) the spectral
distribution $f_{out}(E)$ can be calculated along the same line as
above.

The effect of the induced time-dependent DB potential can be seen in
the plot of $f_{out}(E)$ in Fig.~\ref{coupling}~c), where $e\alpha
V^{\texttt{ac}}=0.1\Delta$. For the adiabatic (not shown) and
intermediate frequencies the main effect is to shift the resonance in
energy, away from $\epsilon_d$, with the shift increasing for
increasing frequency. While the size of the shift depends on $\omega$
and $|\alpha V^{\texttt{ac}}|$, the direction, up or down in energy,
is determined by the sign of $\alpha V^{\texttt{ac}}$. The origin of
the resonance shift can be understood by noting that the discharging
of the DB-region, at low and intermediate frequencies, takes place
during a time interval much shorter than the period ${\mathcal T}$. As
a consequence the electron leaving the DB-region sees an essentially
instantaneous potential $U(t)$. We can thus describe the shift by
considering an effective, time-dependent level energy $\epsilon_d(t)$,
depicted in Fig.~\ref{coupling}~a). For the parameters in
Fig.~\ref{coupling}~c), the effective energy $\epsilon_d(t)$ is slightly
below $\epsilon_d$ when the electron is emitted. The effective level
picture is supported by the fact that the expressions for $f_{out}(E)$
in the adiabatic and intemediate frequency regimes can be found by
substituting $\epsilon_d \rightarrow \epsilon_d(t)$ in
Eq. (\ref{adiabiasfoutwt}) and $\epsilon_d t' \rightarrow \int_{0}^{t'} \epsilon_d(t'')dt''$ in Eq. (\ref{cB})
respectively, where $\epsilon_d(t)=\epsilon_d+e\delta
U(t)$. Importantly, the relatively small magnitude of the shift, given
that $e\alpha V^{\texttt{ac}}=0.1\Delta$, results from the emission
taking place soon after contact B starts to open, \textit{i.e.} close
to $t={\mathcal T}/2$, where the time dependent part $\delta U(t)$ is
small. The lower the frequency, the closer to $t={\mathcal T}/2$ the
emission takes place. We also note that the fraction of the driving
period during which the emission takes place becomes larger when the
frequency increases, explaining the broadening of the resonance with
increasing frequency seen in Fig.~\ref{coupling}~c).

The picture with emission from a time-dependent turnstile level at $\epsilon_d(t)$ breaks
down for frequencies $\hbar\omega\sim\Delta$. In this high frequency regime the main effective of a finite $\delta U(t)$
is to modify the magnitude and width of the individual Floquet fringes. In particular, an asymmetry between fringes corresponding
to absorption and emission of Floquet quanta is clearly visible in Fig.~\ref{coupling}~c).
The details of this asymmetry can be analyzed in terms of which-energy-path arguments, similar to above, 
this is however outside the scope of the present article.

The main conclusion from this analysis is that the electron
distribution, and hence the transferred charge, in the physically most
interesting intermediate frequency regime, is to large extent
unaffected by a time-dependent component of the potential inside the
DB. Only when the time-dependent component becomes comparable to the
level spacing $\Delta$, of the order of $meV$ in closely related
experiments, \cite{Feve,Mahe} are the properties of the turnstile
significantly modified. In our opinion, this investigation provides
strong evidence for the robustness of the turnstile proposed in
Ref. \onlinecite{Battista}.

\section{Conclusions}\label{conclusionsec}
We have performed a detailed theoretical investigation of the spectral
properties of electrons emitted from an on-demand single electron
source. This was done by analyzing a combined single particle
source-spectral detector system implemented with edge states in a
multiterminal conductor. The single particle source and spectrometer
consisted of an electron turnstile and a single-level quantum dot
respectively. The distribution function of the electrons emitted by
the source was investigated via the direct current flowing through the
spectroscopic dot. We investigated the spectral distribution for three
physically distinct frequency regimes; adiabatic, intermediate and
high. It was found that in the adiabatic and intermediate regimes, the
distribution is narrowly peaked around the energy of the turnstile
resonance. At the cross-over to high frequencies the peak splits up,
developing Floquet fringes. At high frequencies an analysis of the
properties of the fringes and their relation to the charge transferred
through the turnstile was examined, highlighting the role of
which-energy-path interference. The robustness of the turnstile
operation in the optimal regime was assessed, providing evidence for
a large resilience to capacitive stray couplings and rectification
effects. Moreover, in the ideal turnstile regime we derived an
expression for the wavefunction of single electrons emitted from the
turnstile and explained how to relate this to the full manybody
wavefunction of the emitted particles. Our findings motivate 
an experimental investigation of the spectral
distribution of electrons emitted from on-demand single electron
sources and put in prospect an observation of Floquet fringes, or
sidebands, directly in the electron distribution.

\section{Acknowledgements}
We acknowledge M. Moskalets, M. B\"uttiker, J. Splettstoesser, M. Albert and C. Flindt for constructive comments 
on an earlier version of the manuscript.
We also acknowledge support from the Swedish VR.

\appendix

\section{Solving the time-dependent Schr\"{o}dinger
  equation}\label{appendixB}
To solve Eq. (\ref{eqsystem}) we first Laplace transform the lower equation
\begin{equation}
i\hbar \big[sc_E(s)-c_E(t=0)]=E c_E(s)+G_d(s)
\label{laplaceeqnsystem}
\end{equation}
where $G_{d}(s)=\int_0^{\infty}e^{-st}J(t)c_{d}(t)\mbox{
}dt$. This directly gives ${c}_E(s)=G_d(s)/(i\hbar
s-E)$ and in the continuum limit (with density of states
$\nu$):
\begin{equation}
 \sum_E c_E(s)=\sum_E\frac{G_d(s)}{i\hbar s-E}\simeq\nu\int\frac{G_d(s)}{i\hbar s-E}\mbox{ }dE.
\end{equation}
Since $G_d(s)$ is energy independent we can perform the energy integral
\begin{eqnarray}
\nu\int\frac{G_d(s)}{i\hbar s-E} dE=\nu G_d(s)(-i\pi).
\end{eqnarray}
The inverse Laplace transform $L^{(-1)}\big[\sum_E c_E(s)\big]=\sum_E c_E(t)=-i\pi\nu J(t)c_d(t)$ 
can be substituted in the first equation of Eq. (\ref{eqsystem}) giving Eq. (\ref{cdt}).

\end{document}